\documentclass[11pt]{article}

\usepackage[letterpaper,margin=1in]{geometry}
\usepackage[utf8]{inputenc}
\usepackage[T1]{fontenc}
\usepackage{amsmath,amssymb}
\usepackage{graphicx}
\usepackage{booktabs}
\usepackage{threeparttable}
\usepackage{longtable}
\usepackage{pdflscape}
\usepackage{float}
\usepackage{caption}
\usepackage{authblk}
\usepackage{setspace}
\usepackage[round]{natbib}
\usepackage{microtype}
\usepackage{xcolor}
\usepackage{array}
\newcolumntype{L}[1]{>{\raggedright\arraybackslash}p{#1}}
\newcolumntype{R}[1]{>{\raggedleft\arraybackslash}p{#1}}
\usepackage[colorlinks=true,citecolor=blue,linkcolor=blue,urlcolor=blue]{hyperref}

\title{\textbf{Catch-Up and Come Home:\\ Economic Convergence and Return Migration}}

\author[1,3]{Kinga Varga}
\author[2]{Bence Koll\'anyi}
\author[3,1,4]{Johannes Wachs\thanks{Corresponding author: \href{mailto:johannes.wachs@uni-corvinus.hu}{johannes.wachs@uni-corvinus.hu}. The authors acknowledge support from the Hungarian Scientific Research Fund (Excellence-149614).}}
\affil[1]{\small ELTE Centre for Economic and Regional Studies}
\affil[2]{\small ELTE Centre for Social Sciences}
\affil[3]{\small Center for Collective Learning, Corvinus Institute of Advanced Studies, Corvinus University of Budapest}
\affil[4]{\small Complexity Science Hub, Vienna}

\date{}

\begin{document}

\maketitle

\begin{abstract}
\noindent The return migration of skilled workers provides origin countries many important benefits. The likelihood of return migration is thought to depend on relative economic opportunities, but quantitative evidence of its relative importance remains limited. Here we measure return migration rates using the public profiles of software developers on GitHub, geocoded across ten waves between 2012 and 2026, in a panel containing more than 270,000 international movers. Within five years of departure, 8.8\% of movers have returned. Origin-country rates range from under 1\% to 16\%. Hazard models with corridor fixed effects show that migrants whose home countries experience greater catch-up growth return at significantly higher rates. Immutable features of country pairs such as cultural or geographic distance do not explain the estimate. We forecast future return rates using IMF growth projections, predicting, for example, that India’s return rate will rise 12\% by 2035 and China’s 7\%. The difference is driven mostly by India’s faster projected growth at home. As developing economies close income gaps with the destinations their emigrants have settled in, they recover more of the workers they lost.
\vspace{0.6em}

\noindent\textbf{Keywords:} return migration; brain drain; brain circulation; high-skilled migration; digital trace data; economic convergence
\end{abstract}

\onehalfspacing
\newpage
\section{Introduction}

Whether sending countries gain or lose from high-skilled emigration depends significantly on how many people come back. Indeed while the advantages for countries receiving high-skilled workers are clear \citep{diodato2022}, recent work emphasizes that sending countries benefit from mobility as well \citep{saxenian2007,batista2025}: high-skilled migrants send remittances \citep{khanna2022abundance}, share information and ideas across borders \citep{agrawal2006, ganguli2015}, and some return home with new skills and knowledge \citep{amanzadeh2024}.

While return migration merits attention as both an outcome of high-skilled mobility and as a potentially consequential factor in development, it is difficult to study because it requires observing individuals across two mobility events rather than one. To count a returning migrant inventor, for instance, one must see her patent in her country of origin, then in the country she moves to, and then in her country of origin again. As a consequence, estimates of the extent of high-skilled return migration are scarce, and we know little about which characteristics of sending countries, or of sending-receiving country pairs, predict variation in return, or indeed whether return rates differ substantially between comparable countries.

However, theory and evidence from specific countries both suggest that economic convergence should predict return migration. In the canonical models of temporary migration, migrants return once the income gap that drew them abroad has narrowed enough, or once the savings and skills they accumulated are worth more at home than another year away would add \citep{dustmann2003, dustmann2016}. Single-country studies find that return responds to economic conditions at home. Return is also thought to be a common outcome, accounting for roughly a quarter to a third of all international migration \citep{azose2019, king2022}, and returnees carry skills, capital, and practices home \citep{wahba2022}. Nonetheless, the effect of convergence on return has not been estimated across many origins and destinations at once. The one large-scale measurement of skilled return reports that 38\% of skilled migrants return within a decade and that return rises with origin income, but it draws on proprietary employment records and estimates a cross-country gravity relationship rather than the within-corridor effect of a closing gap \citep{amanzadeh2024}.

Here we study the dynamics of return migration of high-skilled workers using digital trace data from GitHub, the largest online platform for collaborative software development. Software developers are a high-skill occupation that is globally mobile, widely demanded, and highly visible online. We take ten waves of profile locations between 2012 and 2026, geocode them, and assemble a panel of several million located developers. Of these, 277,502 appear in at least two countries, and 13,670 return to the country in which we first observed them. The panel covers origins and destinations worldwide, spans fourteen years, and resolves migration at the level of individual careers. Five years after leaving, 8.8\% of movers have returned, with origin-level rates ranging from under 1\% for Tunisia and Cuba to 16\% for Spain.

Catch-up growth explains much of this variation. We model the annual hazard of return as a function of how far the income gap between origin and destination has closed since a developer left, identified within corridor, so that two engineers who left India for the United States in different years, and therefore faced different growth paths, are compared with each other. With corridor and year fixed effects, cumulative convergence raises the return hazard (coefficient $0.315$, $p<0.01$). India's income catch-up with the United States between 2015 and 2025, for example, corresponds to about a ten percent increase in the annual rate at which its developers return. The relationship is robust to controls for immutable differences between origin and destination countries such as cultural and geographic distance. We take a conservative approach to measurement throughout: because home is inferred from a developer's first appearance and residence is read from self-reports that update with a lag, both sources of error pull this coefficient toward zero, and correcting either raises it (Section~\ref{sec:robust}).

We use the model to forecast changes in return migration rates by country under IMF growth projections to 2035. Projected changes vary widely across origin countries through the combination of three factors: origin growth, the growth of destinations, and the distribution of emigr\'es across destinations. India's home economy is projected to outgrow the destinations its diaspora occupies by a wide margin, and its developers' return rate rises 12\% in the forecast. China's developers occupy much the same destinations, but their home economy grows more slowly, so their return rises about 7\%. For Japan and other advanced economies, whose destinations are projected to outpace them, return rates decline instead.

We make two contributions. Substantively, we provide global, individual-level evidence that convergence drives the return of high-skilled workers, the mechanism at the center of temporary-migration theory, and we do so for an occupation central to the modern knowledge economy. The return rates we observe are lower bounds, because most movers enter the panel late and self-reported locations update with a lag, so we do not claim that 8.8\% is the rate at which skilled workers return within five years. For example, previous work using employment records, which refresh faster, record higher levels \citep{amanzadeh2024}. The value of an open, individual-level panel lies in its within-corridor variation, which holds the fixed frictions of a corridor constant and lets us estimate the determinants of return and use that estimate in a forecast. Methodologically, we show that public platform data can support demographic measurement and we validate the corridor structure against the best available external estimates of global migration flows \citep{gaskin2026}.

The paper proceeds as follows. Section~\ref{sec:theory} places return migration within the economics of temporary migration and the measurement literature it depends on. Section~\ref{sec:data} describes the panel, the measurement of migration and return, and the covariates. Section~\ref{sec:landscape} reports the descriptive geography of developer migration and return, Section~\ref{sec:results} the hazard estimates, and Section~\ref{sec:robust} their robustness to fixed bilateral frictions. Section~\ref{sec:forecast} projects return rates to 2035 under IMF growth paths, and Section~\ref{sec:conclusion} concludes.

\section{Return migration and its measurement}\label{sec:theory}

The economics of temporary migration treats return as a decision about timing. A migrant abroad weighs the wage premium at the destination against a preference for life at home, and in the canonical models returns once the income gap has narrowed enough, or once the savings and skills accumulated abroad are worth more at home than another year away would add \citep{dustmann2002, dustmann2003, dustmann2016}. For some migrants return is the plan from the outset, the years abroad undertaken to finance investment or consumption back home \citep{mesnard2004, bossavie2025}.

These models suggest that return should rise as origin incomes catch up with destination incomes. The claim concerns the income gap closing over time, not its level at any single moment, so the natural test compares migrants who left the same origin for the same destination at different points along the origin's growth path, holding constant everything permanent about the route. Read this way, convergence is a prediction about the timing of return within a corridor, and it is separate from the question of which corridors show high return overall.

The role of selection into migration makes it difficult to test these ideas with aggregate data. Specifically, who emigrates and who returns are both selected groups of a population \citep{borjas1996, wahba2015}. Highly skilled emigrant workers are positively selected: they are more educated than the non-migrant population of their country of origin \citep{docquier2007brain,grogger2011income}. The magnitude of this selection varies by origin, as it depends on the skill-related earnings difference between origin and destination rather than on the average income gap alone \citep{grogger2011income}, with the highest-scoring graduates of the Indian Institutes of Technology emigrating at the highest rates \citep{choudhury2023}. When the composition of movers and returnees shifts with economic conditions, the net flow between two countries mixes the return decision together with who was abroad in the first place. The convergence mechanism would likely be reflected in the duration of individual spells abroad rather than in the balance of flows, which is why data on individual migration events is valuable.

Direct evidence for our prediction that convergence predicts return migration exists, but almost entirely from single origins or destinations. Filipino migrants time their return to exchange-rate movements \citep{yang2006}. Mexican migrants to the US react to local opportunity \citep{lindstrom1996}, and immigrants in the Netherlands react to labor market conditions \citep{bijwaardschluter2014}. Of course income and jobs are not the only consideration: the return of Pacific high-achievers depends on family and lifestyle factors, too \citep{gibson2011}, and a migrant's broader intention to return is closely tied to whether their initial expectations of the move have actually been fulfilled \citep{hu2026role}. We are not aware of tests of the same income convergence across many corridors at once, the test that would show whether convergence drives return in general or only in particular places.

How many migrants return matters because returnees contribute substantially to the economies they rejoin. In science, scholars returning under mandated-return programs diffuse knowledge home \citep{kahn2016}, returning African scientists raise their colleagues' output \citep{fry2023}, and returnee scientists rebuild research capacity in their home institutions \citep{fryganguli2026}. In the corporate sector, returnee managers raise local patenting inside multinationals \citep{choudhury2016} and returnee directors raise the valuation of Chinese firms \citep{giannetti2015}; in public life, returnees make more effective local leaders \citep{mercier2016}, and at the level of the whole economy return raises origin-country development outright \citep{bucheli2025}. Some of the gain arrives before anyone moves back, through the knowledge and capital a diaspora channels home \citep{agrawal2011, beine2008, docquier2012}.

This work estimates the effect of a given returnee and takes the number of returnees as fixed. The aggregate value of the channel, however, is the effect per returnee multiplied by the number who return, so it depends as much on the rate of return and its determinants as on the individual effects these studies measure. That rate, and what moves it, is what we estimate here.

Measuring that rate is the standing difficulty, because a return migrant must be observed three times: at home, then abroad, then at home again. Population registers rarely record all three, and demographers have begun to close the gap with digital traces; the most recent review of these measurement problems ends by calling for precisely this, digital footprints that can predict return, track its patterns, and assess its local impacts \citep{wahba2026}. Email logins, social-media profiles, and professional networks have each been used to estimate migration where official sources are silent \citep{zagheni2012, state2014}, and \citet{fiorio2021} show how the timing structure of trace data governs what can be inferred from it. At the aggregate level, model-based methods triangulate global flows from census stocks \citep{abel2014, azose2019}, most recently with machine learning \citep{gaskin2026}.

Software developers are a particularly tractable case for this approach. Platform data have already documented crisis-driven exodus among them \citep{wachs2023}, and proprietary employment histories have produced the first large-scale return rates for skilled workers \citep{amanzadeh2024}. We build on each of these literatures, using open data on this single, well-defined occupation to estimate individual return durations across thousands of corridors and to test the convergence mechanism directly.

\section{Data: a global location panel of software developers}\label{sec:data}

\subsection{Source and construction}

Our data come from GitHub, the platform on which most of the world's open-source software is written and a large share of commercial development is coordinated. GitHub passed 100 million registered developers in early 2023 and reports more than 180 million as of 2025,\footnote{GitHub Blog, 25 January 2023; GitHub Octoverse, October 2025.} and each account carries a public profile with an optional free-text location field that users can set and update themselves. Developers keep these profiles current for professional reasons, because a GitHub page serves as a portfolio shown to employers, collaborators, and recruiters. The resulting geography is informative: the spatial distribution of GitHub activity closely tracks the geography of the software industry measured by conventional sources \citep{wachs2022}, and GitHub records now underpin measures of software investment consistent with national accounts \citep{korkmaz2024} and of the software complexity of national economies \citep{juhasz2026}.

We assemble ten location waves from these public profiles, taken in 2012, annually from 2015 to 2019, in 2021, and annually from 2024 to 2026. The waves combine two sources. Those from 2012 through 2021 come from GHTorrent, the public research archive that mirrored GitHub's metadata for over a decade and became the standard source for large-scale studies of the platform \citep{gousios2012, gousios2013}. GHTorrent stopped updating in 2021, which is why the panel holds no 2022 or 2023 wave. From 2024 onward we collect the waves ourselves by querying the GitHub REST API, which returns the same public fields the archive recorded, so the two sources align. The 2025 wave is itself a composite of two rounds of data collections, conducted in January and March 2025; the second was united with the first in a later revision of the panel, which raised the number of located developers in that wave from 4.04 million to 6.15 million, a 52 percent increase. 

The raw panel covers 44 million accounts and 113 million person-wave observations. Location strings are geocoded to countries through a curated dictionary of 1.2 million distinct strings. To gauge that step we ran a blind audit of 1,000 randomly sampled strings: 97.6\% of strings and 99.7\% of users are assigned correctly, and the residual errors are non-place strings rather than misassigned places. After removing bots and organizations and accounts without a resolvable location, the analysis panel holds 42.6 million located person-waves for 10.0 million developers. Appendix~\ref{app:pipeline} documents the pipeline and the audit.

Data of this kind are emerging sources in economics and demography. \citet{nagle2019} uses open-source activity to estimate firm productivity and \citet{petralia2025} studies open-source software as a platform for firm innovation, while \citet{wright2023} show that country-level GitHub participation predicts the founding of innovative ventures. Closer to our setting, a growing literature uses geolocated developers as a sampling frame for the global programming workforce. Much of it measures how geography shapes collaboration and productivity: colocation effects in collaboration networks \citep{goldbeck2025}, border effects in cross-country collaboration \citep{abouelkomboz2024el}, and agglomeration spillovers in productivity \citep{abouelkomboz2024cesifo}. The same data capture labor-market behavior as well: contribution activity rises when developers search for jobs \citep{abouelkomboz2025}, and more visible contribution records move workers toward larger firms \citep{gupta2024}. Closest to our concerns, \citet{birkholz2026} model the global software production network from millions of geolocated users and find sorting patterns suggestive of brain drain, while \citet{wachs2023} uses self-reported locations to track developer relocation during the Russian invasion of Ukraine. 

\subsection{Measuring migration and return}

We define a mover as a developer observed in at least two countries at different times. Our panel contains 277,502 of such movers, spread across more than 8,100 origin-destination corridors; of these, developers from 196 origin countries with a defined convergence measure enter our main estimation. We define home as a developer’s first observed country, and a return as a subsequent reappearance in that country after living abroad. Within our observation window, 13,670 movers make such a return.


Each mover enters the risk set in the first wave after departure; the departure wave can contain neither a return nor any convergence, so it is excluded by construction. Movers are not censored when they move onward, so the estimand is the hazard of returning home from anywhere abroad. 

The resulting risk set comprises 547,619 person-periods with a 2.50\% per-period event rate; movers with no observed wave after departure, most of them 2026 departures, contribute no person-periods, so 215,305 movers enter the risk set. Table~\ref{tab:descriptives} summarizes the panel.

\begin{table}[htbp]
\centering
\small
\begin{tabular}{lr}
\toprule
Location waves & 10 (2012--2026) \\
Movers (observed in $\geq$2 countries) & 277,502 \\
Returnees & 13,670 \\
Origin--destination corridors & 8,114 \\
Person-periods at risk & 547,619 \\
Per-period return rate & 2.50\% \\
Median years to return & 3 \\
Share of movers departing 2024 or later & 66.6\% \\
\bottomrule
\end{tabular}
\caption{Composition of the developer location panel. A mover is a developer observed in at least two countries; a returnee is a mover later seen again in the first country observed. For example, a developer first located in Bangalore, then Berlin, then Bangalore is one mover and one returnee.}
\label{tab:descriptives}
\end{table}

\subsection{Observation windows and external validity}

Pooled return rates understate actual return rates, because two thirds of movers departed in 2024 or later and have had at most two years to return. Reported by departure cohort, the rate is 8.8\% within five years for the cohorts we can follow that long, and it continues to rise for the cohorts we observe longer. Observed rates are therefore lower bounds on lifetime circulation, which is why the models below compare movers within cohort and corridor rather than relying on levels. Employment-history data, which refresh faster than profiles, record higher levels on a different denominator \citep{amanzadeh2024}; within-panel comparisons are unaffected by the difference.

We benchmark the corridor structure against the machine-learning flow estimates of \citet{gaskin2026}. Across all 48{,}620 corridors among the countries our panel covers, most of which contain no movers in our data, departures correlate with their predicted bilateral flows at a Spearman rank of 0.49 (Pearson 0.60 on log-transformed variables). The two populations differ (we have developers rather than all migrants), so we read this as strong co-movement rather than level validation.

\subsection{Covariates}\label{sec:covariates}

Our key regressor is cumulative economic convergence since departure. For a mover who left origin $o$ for destination $d$ in year $t_0$, convergence in year $t$ is
\begin{equation}
\mathrm{EC}_{od,t_0}(t) = \bigl(\ln y_{o,t} - \ln y_{d,t}\bigr) - \bigl(\ln y_{o,t_0} - \ln y_{d,t_0}\bigr),
\end{equation}
where $y_{c,t}$ is GDP per capita at purchasing power parity in country $c$ and year $t$. The measure is zero in the departure year and turns positive once the origin has closed part of its log income gap with the destination. A positive value means the origin grew faster than the destination after departure; a value near zero means parallel growth; a negative value means the gap widened. Income series are GDP per capita, PPP, in current international dollars, from the World Bank World Development Indicators (series NY.GDP.PCAP.PP.CD) through 2024, extended to 2025 with IMF estimates. Because origin and destination are compared within the same year before differencing, the common international price level drops out. IMF projections enter only the forecast of Section~\ref{sec:forecast}, so estimation rests on the historical series alone. Taiwan, which the World Bank does not cover, takes its GDP per capita from the IMF World Economic Outlook on the same purchasing-power basis.

Economic convergence has theoretical grounding and empirical support. Whether its effect on return is genuine, or merely reflects the immutable frictions it correlates with, such as the cultural and geographic distance between origin and destination, has not been tested at a global scale. To separate the two we collected several bilateral features that offer alternative explanations: Kogut--Singh cultural distance over Hofstede dimensions \citep{kogut1988, hofstede2001}, physical distance, time-zone offset, and the destination's integration policy, measured by the Migrant Integration Policy Index (MIPEX) \citep{solano2020}, which scores countries on how far their laws grant immigrants secure residence, family reunification, and access to the labor market.

\section{The global landscape of developer return migration}\label{sec:landscape}

The geography of developer migration follows the geography of the global technology industry. The United States is by far the largest destination, receiving close to one in five of all movers, more than the next two destinations combined. Germany has recently become the second-largest destination, ahead of the United Kingdom, with Canada and the Netherlands next. On the sending side, India is the largest origin by a wide margin, ahead of the United States, Russia, the United Kingdom, and China. Migration is concentrated at both ends: the five largest destinations absorb just under half of all movers, and the largest single corridor, from India to the United States, is three times the size of the next.

\begin{figure}[htbp]
\centering
\includegraphics[width=0.9\linewidth]{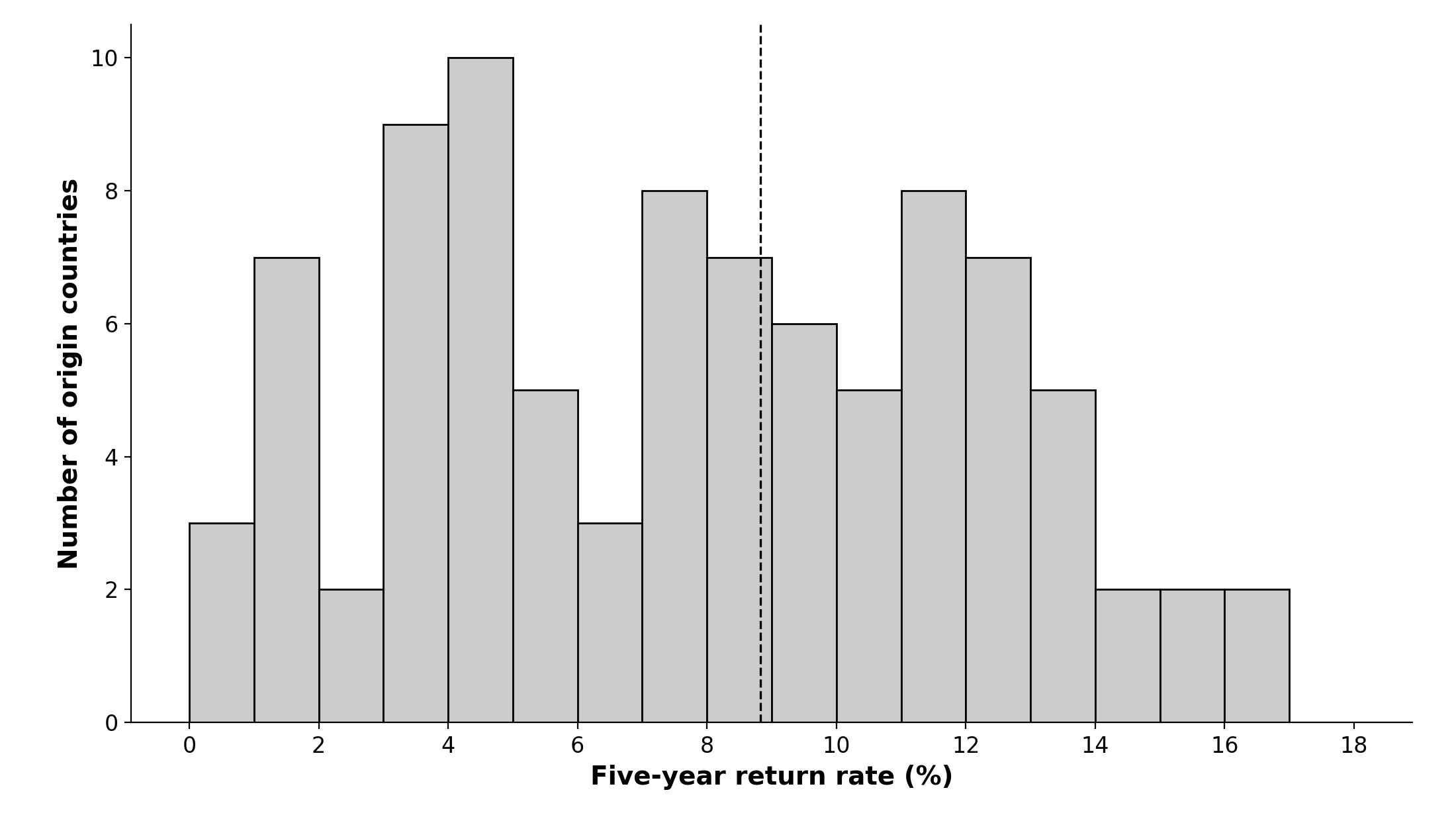}
\caption{Distribution of five-year return rates across origin countries. For each origin we take the departure cohorts that an observed wave follows exactly five years later (2016, 2019, and 2021) and compute the share that returned home within five years; the histogram pools these origin-level rates across the 91 origins with at least 50 such movers. The dashed line marks the global rate of 8.8\%. Rates run from under 1\% (Tunisia, Cuba) to 16\% (Spain), with the post-communist EU members near the top and the non-EU east near the bottom.}
\label{fig:hist}
\end{figure}

Return rates, by contrast, vary widely across origins: five years after departure, the share of an origin's movers who have returned ranges from under 1\% to 16\%, a more than twentyfold spread (Figure~\ref{fig:hist}). Among origins with at least 1,000 movers, Spain has the highest rate, at 16\% within five years; India, the largest sender, recovers under 5\%, while Tunisia, Cuba, and Venezuela are near 1\%. Return is most common where income gaps are smallest and rarest where they are largest, the pattern the convergence hypothesis predicts. Not all of the variation follows income, however: the post-communist members of the European Union return at nearly Western rates, Poland at 15\% and the Czech Republic higher still, plausibly because free movement makes return easy to attempt and easy to reverse, while the non-EU east is near the bottom of the same distribution. The rate bears no relation to how many developers a country sends: the largest senders span almost the whole range. Appendix~\ref{app:supp} reports this five-year rate for all origins with at least 1,000 movers, alongside the pooled rate across all departure cohorts.

Absolute returns are concentrated in the largest diasporas. India's is the largest by a wide margin, about 33,000 developers abroad at the 2026 baseline, and only a small share has returned so far, so even a modest rise in its return rate would bring back many developers. Whether economic convergence raises return rates, and by how much, is what we estimate next.

\section{Methods and Results}\label{sec:results}

\subsection{Empirical approach}

We model return with a discrete-time hazard and a logit link \citep{allison1982}. The hazard that mover $i$, who left origin $o$ for destination $d$ in year $t_0$, returns in wave $t$ is
\begin{equation}\label{eq:hazard}
h_{it} = \Lambda\bigl(\beta\,\mathrm{EC}_{od,t_0}(t) + \delta\,\tau_{it} + \gamma_{od} + \lambda_{t}\bigr),
\end{equation}
where $\Lambda$ is the logistic function, $\tau_{it}$ counts years since departure and enters linearly, $\lambda_t$ are wave-year effects, and $\gamma_{od}$ are corridor fixed effects in the most demanding specification; the intermediate columns replace $\gamma_{od}$ with origin and destination country effects. The identifying variation narrows across columns: under corridor fixed effects the convergence coefficient is identified purely by comparing movers on the same corridor who departed at different points along the origin's growth path. Standard errors are clustered two-way, by origin and by destination country \citep{cameron2011}. Estimation drops the 12,154 person-periods whose convergence measure is undefined because the origin or destination lacks a continuous GDP per capita series, many of them Venezuelan corridors, and, under corridor fixed effects, the corridors in which the outcome never varies, almost all of them corridors with no observed return; 490,756 person-periods remain.\footnote{A linear probability model comes to the same conclusion that a unit of convergence raises the return probability by $0.011$ (SE $0.003$, $N = 534{,}676$).}

The sign of the convergence coefficient depends on the control set (Table~\ref{tab:ecladder}). Without country controls the estimate is negative because corridors with large income gaps and fast convergence, for example India to the United States, also have persistently low return rates, and these cross-corridor differences dominate the pooled estimate. With origin and destination effects the coefficient is positive, and with corridor fixed effects, which compare only movers who took the same corridor, it is $+0.315$ (SE 0.098, $p<0.01$). Return propensity falls with time abroad throughout, the standard duration pattern.

\begin{table}[htbp]
\centering
\begin{tabular}{lcccc}
\toprule
 & \multicolumn{4}{c}{DV: Return event} \\
\cmidrule(lr){2-5}
 & (1) & (2) & (3) & (4) \\
\midrule
Cumulative economic convergence & $-1.193^{+}$ & $0.110$ & $0.180^{+}$ & $0.315^{**}$ \\
 & (0.633) & (0.301) & (0.099) & (0.098) \\
\addlinespace
Years since departure & $-0.103^{***}$ & $-0.130^{***}$ & $-0.124^{***}$ & $-0.119^{***}$ \\
 & (0.016) & (0.009) & (0.009) & (0.009) \\
\midrule
Year FE & \checkmark & \checkmark & \checkmark & \checkmark \\
Origin country FE & & \checkmark & \checkmark & \\
Destination country FE & & & \checkmark & \\
Corridor FE & & & & \checkmark \\
\midrule
Observations & 491,545 & 491,479 & 491,405 & 490,756 \\
Base event rate & 0.0250 & 0.0250 & 0.0250 & 0.0250 \\
\bottomrule
\end{tabular}
\caption{Economic convergence and the return hazard. Discrete-time logit hazard models on the post-departure risk set (per-period base rate 2.50\%), consistent sample across columns. Two-way clustered standard errors (origin and destination country) in parentheses. $^{+}\,p<0.1$, $^{*}\,p<0.05$, $^{**}\,p<0.01$, $^{***}\,p<0.001$.}
\label{tab:ecladder}
\end{table}

\begin{figure}[htbp]
\centering
\includegraphics[width=0.95\linewidth]{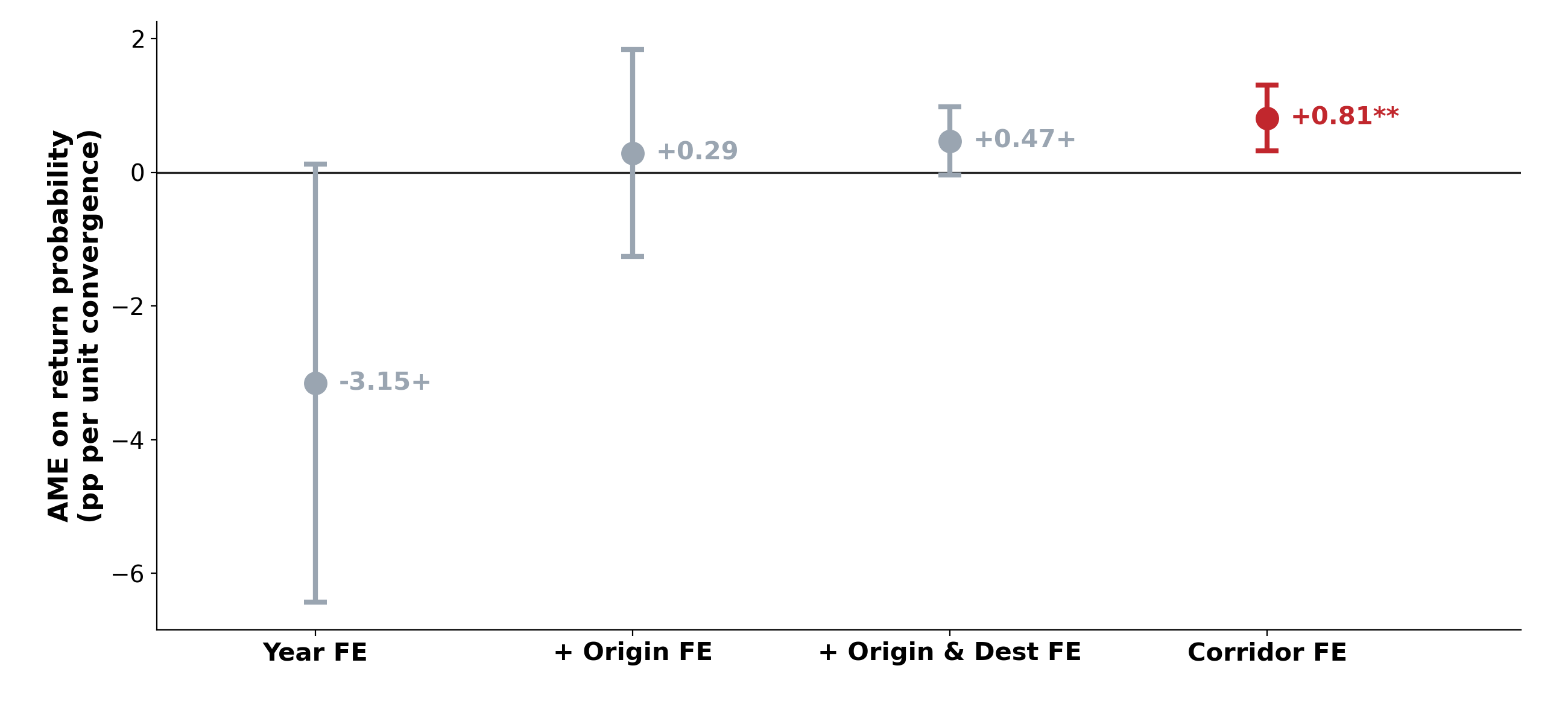}
\caption{Average marginal effect of a unit of cumulative convergence on the per-period return probability, across specifications, with 95\% confidence intervals. Discrete-time logit hazard, return event on cumulative convergence and years since departure with fixed effects as labeled, on the consistent sample estimable under the corridor specification; AME $= \beta \times$ mean $\hat{p}(1-\hat{p})$ with fixed effects in the fitted probabilities; two-way clustered standard errors by origin and destination. $+\,p<0.1$, $^{*}p<0.05$, $^{**}p<0.01$, $^{***}p<0.001$.}
\label{fig:ame}
\end{figure}

The magnitude is most readily interpreted in probability terms (Figure~\ref{fig:ame}). In the corridor specification, one unit of cumulative convergence raises the return probability by about 0.81 percentage points, against a base rate of 2.5\%. A full unit is large: it would require the origin economy to grow roughly 2.7 times as fast as the destination over the entire period abroad, which few corridors approach. India's catch-up with the United States between 2015 and 2025 was a more representative $+0.35$, corresponding to approximately a ten percent increase in the return rate. Applied across hundreds of thousands of diaspora developers, an effect of this size is consequential, as the following section shows.

\section{Distance, culture, and policy}\label{sec:robust}

Convergence varies from year to year, but the corridors on which we measure it also differ in more permanent ways. The physical and cultural distance between an origin and a destination could each shape return, and either could account for the convergence estimate if it happens to correlate with which corridors grow. This section separates convergence from these fixed features of a corridor, and then from destination integration policy.

Geographic distance does not account for the convergence effect: log distance and time-zone offset do not predict return once country composition is held fixed (Appendix~\ref{app:supp}). Developers who moved to distant destinations return at about the same rate as those who moved to nearby ones.

Developers from culturally distant origins return less often: the coefficient on Kogut--Singh cultural distance is $-0.069$ ($p<0.001$). This does not account for the convergence effect, because culture and convergence are different in kind (Table~\ref{tab:horserace}). Convergence remains positive and significant whether cultural and geographic distance enter as controls or corridor fixed effects absorb every fixed bilateral attribute, and it varies year by year along the origin's growth path while cultural distance is fixed. We cannot say from these data why culture dampens return, whether through deeper integration in culturally similar destinations, adaptation over time, or selection into who leaves for a distant country in the first place, so we report the pattern without adjudicating among these channels. One natural follow-up, that catch-up might pull less strongly where culture is distant, does not hold: the convergence-by-culture interaction is small and insignificant (Appendix~\ref{app:interaction}).

\begin{table}[htbp]
\centering
\begin{tabular}{lcccc}
\toprule
 & \multicolumn{4}{c}{DV: Return event} \\
\cmidrule(lr){2-5}
 & (1) & (2) & (3) & (4) \\
\midrule
Cumulative economic convergence & $-0.812^{+}$ & $0.352$ & $0.255^{**}$ & $0.292^{**}$ \\
 & (0.437) & (0.242) & (0.086) & (0.097) \\
\addlinespace
Cultural distance (Kogut--Singh) & $-0.065^{*}$ & $-0.042^{*}$ & $-0.065^{***}$ & --- \\
 & (0.025) & (0.021) & (0.016) & \\
\addlinespace
Log distance (km) & $0.005$ & $0.077^{+}$ & $0.062^{+}$ & --- \\
 & (0.060) & (0.046) & (0.035) & \\
\addlinespace
Timezone difference (hours) & $-0.018$ & $-0.034$ & $-0.020$ & --- \\
 & (0.028) & (0.021) & (0.015) & \\
\addlinespace
Years since departure & $-0.102^{***}$ & $-0.128^{***}$ & $-0.126^{***}$ & $-0.121^{***}$ \\
 & (0.014) & (0.010) & (0.010) & (0.010) \\
\midrule
Year FE & \checkmark & \checkmark & \checkmark & \checkmark \\
Origin country FE & & \checkmark & \checkmark & \\
Destination country FE & & & \checkmark & \\
Corridor FE & & & & \checkmark \\
\midrule
Observations & 508,846 & 508,845 & 508,845 & 473,199 \\
Base event rate & 0.0250 & 0.0250 & 0.0250 & 0.0250 \\
\bottomrule
\end{tabular}
\caption{Horserace: economic convergence against fixed bilateral frictions. Discrete-time logit hazard models on the post-departure risk set. Cultural distance, geographic distance, and time-zone difference are corridor-invariant and are absorbed by the corridor fixed effects in column~(4). Two-way clustered standard errors (origin and destination country) in parentheses. $^{+}\,p<0.1$, $^{*}\,p<0.05$, $^{**}\,p<0.01$, $^{***}\,p<0.001$.}
\label{tab:horserace}
\end{table}

Destination integration policy does not account for the effect either. A destination that integrates its immigrants more fully, through secure residence, family reunification, and access to the labor market, might hold onto them even as their home economies grow, muting the convergence channel. We find no such moderation: where MIPEX scores are available, adding the destination's policy level leaves the convergence effect unchanged, the level itself does not predict return, and the two do not interact. Convergence pulls developers home from restrictive and welcoming destinations alike.

Convergence remains positive and significant when we measure income in constant-PPP dollars, which strip out price-level drift, in place of current dollars. We keep current international dollars as the headline because that series runs through 2025, matches the IMF projections behind the forecast, and is the purchasing-power comparison a developer weighing return faces. A remaining question is whether GDP per capita, an economy-wide average, reflects what software developers earn; it does, since developer pay tracks GDP per capita at PPP closely across countries (Appendix~\ref{app:salary}).

\subsection{Convergence estimates and alternative subsets}\label{sec:lowerbound}

Two features of the data suggest our estimates of the effect of convergence on return migration are lower bounds. We assign each developer's home from the first country in which she appears, so a developer who had already emigrated when the panel opens is placed on a corridor she never took, her destination mistaken for her origin. Locations are also self-reported and lag real moves, so a return home can go unrecorded when a developer stops maintaining her profile. Both errors work against finding an effect, so the full-sample coefficient we report is likely an underestimate of the effect of convergence on return migration. 

First we consider that some users may have first migrated before the start of our dataset. This would cause us to count actual return migrations as first time migrations. One way to test this in our data is to identify the likelihood a user is from the first country we record them in using their name. The display name is where users typically list their ``real'' names.  A statistical name-to-nationality service predicts likely nationalities from the profile's display name, and we re-estimate on movers whose predicted nationalities include the home we assigned. Dropping movers whose names contradict their coded origin raises the convergence coefficient to $+0.424$ ($p<0.01$), about 35\% above the headline. The second uses the panel: among the 86\% of movers observed at home in at least two waves before leaving, whose origin therefore cannot be a left-censoring artifact, the coefficient is $+0.301$ ($p<0.01$), close to the headline. Neither correction lowers the estimate (Appendix~\ref{app:names}).

The stricter name cuts lose precision as the sample shrinks. The name test also misjudges developers whose family name comes from a country other than their home: a developer correctly placed in the United States but with an Indian family name is predicted to be Indian and dropped. The matched sample is therefore selected rather than random. Within these limits, a misassigned origin lowers the measured convergence effect.

Second, we consider the bias introduced by users who move then stop updating their profiles. A developer who returns home but does not update her profile is recorded as still abroad, so her return never enters the data and the convergence coefficient is biased downward. We identify the movers whose profiles are current from a signal independent of the location field: an edit, after departure, to the display name, company, blog, or linked accounts fields. A developer who changes these fields is keeping her profile current, so a move home would also appear in her recorded location. Re-estimating the corridor model on these movers raises the convergence coefficient to $+0.448$ ($p<0.05$), about 42\% above the headline (Appendix~\ref{app:names}).

That the coefficient rises, rather than remaining unchanged, is itself informative: if stale profiles hid returns unrelated to convergence, restricting to current profiles would leave the estimate where it was, so the missed returns must fall disproportionately on the corridors where convergence is strongest. Developers who edit their profiles are more recently active, and a returnee is somewhat easier to observe than a developer who remains abroad, so part of the increase reflects which movers stay visible, and not the correction of missed returns alone. Stale profiles lower the measured convergence effect.

\section{The convergence channel to 2035}\label{sec:forecast}

The forecast isolates a single channel: if economies converge as the IMF expects, how does return migration change? The projection takes each corridor's 2026 baseline hazard and shifts it by $\exp(\beta\,\Delta)$, where $\Delta$ is the additional projected convergence on the corridor and $\beta = 0.315$ comes from the corridor specification. Convergence is built from the same income series the model is estimated on, GDP per capita at purchasing power parity from the IMF World Economic Outlook of April 2026, carried to 2035 at each country's terminal projected growth rate. Origin-level figures weight each corridor by its current diaspora stock and 2026 baseline hazard, and the table reports the underlying growth rates in real per-capita terms. Because one pooled coefficient is applied to every corridor, the projection averages over corridors that may each respond differently, so relative changes are more informative than levels. An origin's expected return depends on three factors: how fast its own economy grows, where its diaspora is located, and how fast those destinations grow.

\subsection{Growth and return}
Because the forecast relies on a single model coefficient, it is useful to first establish the magnitude it implies. Consider a country whose income gap with its diaspora's destinations closes one percentage point per year faster than it otherwise would, sustained over a decade. This adds 0.10 log points of cumulative convergence, which corresponds in our model to a $\exp(0.315 \times 0.10) - 1 \approx 3.2\%$ increase in the annual return rate, or about 0.08 percentage points on the 2.5\% base. A differential of half a percentage point per year corresponds to roughly 1.6\% more return, and a full point of slower catch-up to a reduction of about 3.1\%. But this calculation assumes the same convergence premium of the origin country versus its destination pairs: in reality, it is likely that observed convergence will depend significantly on where a country's developers go, which is significantly determined by geography and history.

India and China illustrate how these factors interact. Both send most of their developers to the United States and other high-income economies, so their diasporas are in destinations projected to grow at similar rates. That said, compared to China, India has a relative bias towards the United Kingdom for clear historical reasons, and the UK is forecast to more slowly than the US. Another point of distinction are their home economies. India is projected to grow about a percentage point and a half per year faster than China, closing more of its income gap with those shared destinations by 2035. Our model's projection yields 11.9\% more return for India against 6.8\% for China. 

\begin{figure}[htbp]
\centering
\includegraphics[width=0.82\linewidth]{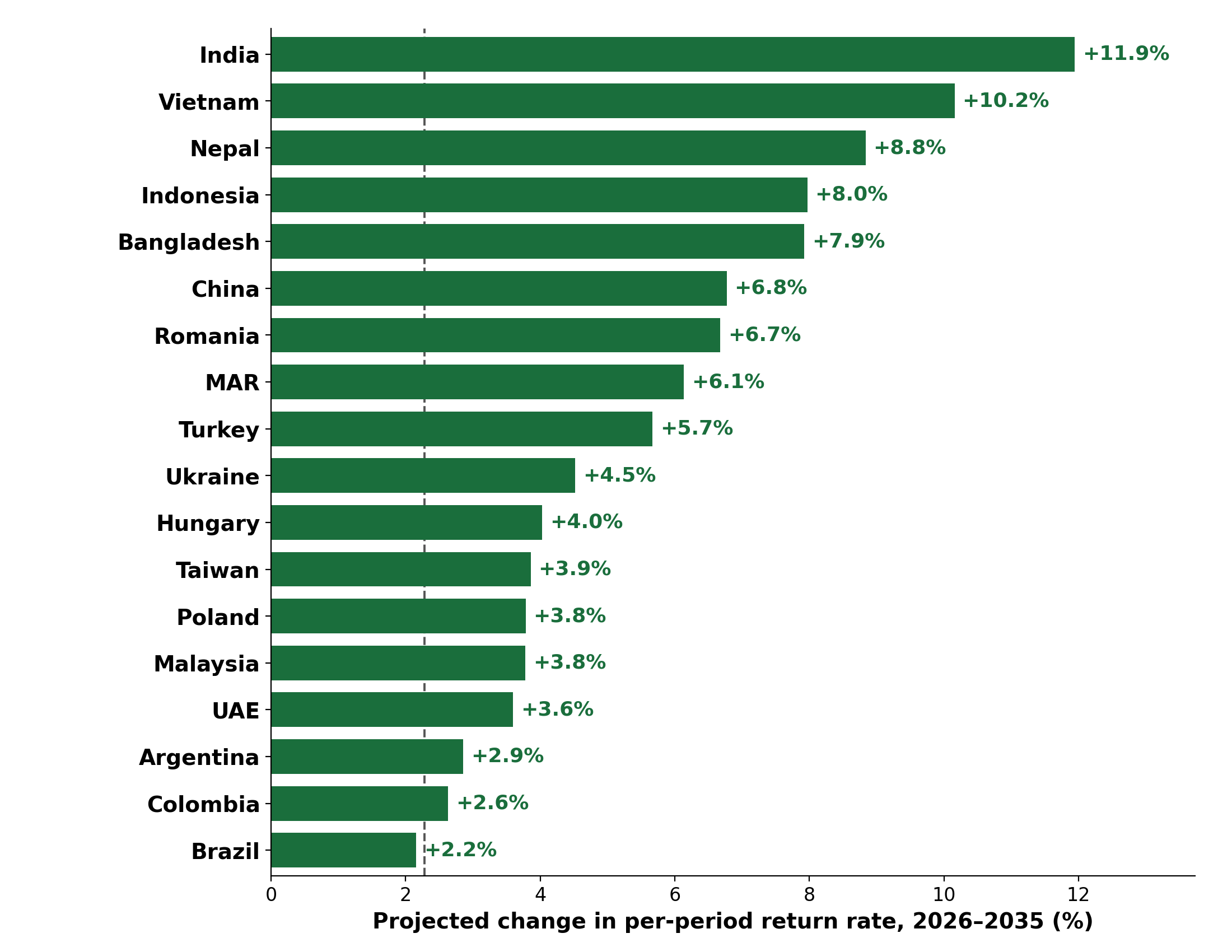}
\caption{Projected percent change in the per-period return rate by origin, 2026 to 2035, under the convergence channel, for the eighteen origins with the largest projected gains (origins with at least 1,000 movers). The dashed line marks the stock-weighted all-origin average of $+2.3\%$. The projection method is described in the text.}
\label{fig:forecastpct}
\end{figure}

In many cases, the third factor, destination growth, drives the projected declines: return falls for Japan ($-1.8\%$) and South Africa ($-2.9\%$), whose developers are in destinations projected to grow faster than the home economy, so the income gap widens rather than closes. Stock-weighted across all origins, the increase is 2.3\% (Table~\ref{tab:forecast2035}, Figure~\ref{fig:forecastpct}). The largest gains accrue to developing economies with large developer diasporas abroad: India ($+11.9\%$), Vietnam ($+10.2\%$), Nepal ($+8.8\%$), Indonesia, and Bangladesh.
\clearpage
{\scriptsize
\renewcommand{\arraystretch}{0.78}
\setlength{\LTleft}{\fill}\setlength{\LTright}{\fill}
\begin{longtable}{L{3.0cm} R{1.4cm} R{1.6cm} R{1.9cm} R{1.6cm} R{1.9cm} R{1.6cm}}
\toprule
Origin & 2026 return rate (\%) & Home growth (\%/yr) & Dest.\ growth (\%/yr) & Gap closed (log points) & Rate change (\%) & Rate change (pp) \\
\midrule
\endfirsthead
\toprule
Origin & 2026 return rate (\%) & Home growth (\%/yr) & Dest.\ growth (\%/yr) & Gap closed (log points) & Rate change (\%) & Rate change (pp) \\
\midrule
\endhead
\midrule
\multicolumn{7}{r}{\footnotesize\itshape continued on next page}\\
\endfoot
\bottomrule
\caption{Forecast change in return migration rates by origin, 2026--2035, under the convergence channel with $\beta = 0.315$. Projected catch-up raises India's return rate by 11.9\%; where destinations are projected to outgrow the origin, return declines. Home and destination growth are real GDP per capita from the IMF World Economic Outlook, April 2026, the destination figure diaspora-weighted; origins with at least 1{,}000 movers, ordered by diaspora size.}\label{tab:forecast2035}\\
\endlastfoot
India & 1.2 & 5.7 & 1.6 & +0.36 & +11.9 & +0.14 \\
United States & 2.4 & 1.8 & 2.0 & -0.03 & -0.8 & -0.02 \\
Russia & 1.1 & 1.3 & 2.4 & -0.14 & -4.2 & -0.04 \\
United Kingdom & 1.5 & 1.0 & 1.7 & -0.05 & -1.5 & -0.02 \\
China & 1.8 & 4.2 & 1.6 & +0.21 & +6.8 & +0.12 \\
Brazil & 2.5 & 2.0 & 1.4 & +0.07 & +2.2 & +0.05 \\
Germany & 2.4 & 1.0 & 1.5 & -0.04 & -1.4 & -0.03 \\
Canada & 1.7 & 1.2 & 1.8 & -0.05 & -1.6 & -0.03 \\
France & 3.3 & 0.8 & 1.4 & -0.05 & -1.5 & -0.05 \\
Ukraine & 0.7 & 2.5 & 1.8 & +0.14 & +4.5 & +0.03 \\
Turkey & 1.5 & 3.3 & 1.4 & +0.17 & +5.7 & +0.08 \\
Spain & 3.3 & 1.0 & 1.4 & -0.02 & -0.6 & -0.02 \\
Italy & 2.6 & 0.8 & 1.2 & -0.02 & -0.7 & -0.02 \\
Belarus & 0.7 & 1.5 & 2.6 & -0.10 & -3.2 & -0.02 \\
Australia & 2.3 & 0.8 & 1.9 & -0.06 & -1.9 & -0.04 \\
Pakistan & 1.4 & 2.2 & 1.7 & +0.04 & +1.3 & +0.02 \\
Netherlands & 1.8 & 0.9 & 1.5 & -0.05 & -1.6 & -0.03 \\
Iran & 1.3 & -0.5 & 1.5 & -0.08 & -2.5 & -0.03 \\
Bangladesh & 1.3 & 4.1 & 1.6 & +0.24 & +7.9 & +0.10 \\
Nigeria & 0.9 & 2.1 & 1.3 & +0.05 & +1.7 & +0.02 \\
Japan & 3.0 & 1.2 & 2.1 & -0.06 & -1.8 & -0.05 \\
Argentina & 1.9 & 2.6 & 1.3 & +0.09 & +2.9 & +0.05 \\
Poland & 3.3 & 2.9 & 1.4 & +0.12 & +3.8 & +0.13 \\
Singapore & 1.3 & 2.0 & 2.1 & 0.00 & 0.0 & 0.00 \\
Sweden & 2.2 & 1.3 & 1.5 & -0.01 & -0.2 & 0.00 \\
Switzerland & 1.9 & 0.8 & 1.5 & -0.03 & -0.9 & -0.02 \\
South Korea & 2.6 & 2.1 & 1.7 & +0.06 & +1.8 & +0.05 \\
Ireland & 1.7 & 1.6 & 1.5 & +0.02 & +0.5 & +0.01 \\
Mexico & 1.5 & 1.3 & 1.6 & -0.01 & -0.2 & 0.00 \\
Egypt & 1.4 & 2.7 & 1.7 & +0.03 & +1.1 & +0.01 \\
Portugal & 3.0 & 2.0 & 1.3 & +0.06 & +2.0 & +0.06 \\
Hong Kong & 1.1 & 1.7 & 2.0 & -0.03 & -1.1 & -0.01 \\
New Zealand & 1.4 & 1.7 & 1.3 & 0.00 & 0.0 & 0.00 \\
South Africa & 1.0 & 0.1 & 1.3 & -0.09 & -2.9 & -0.03 \\
Taiwan & 1.7 & 3.5 & 1.7 & +0.12 & +3.9 & +0.06 \\
Colombia & 2.0 & 2.1 & 1.4 & +0.08 & +2.6 & +0.05 \\
Nepal & 0.7 & 4.7 & 1.6 & +0.27 & +8.8 & +0.06 \\
Greece & 1.8 & 1.8 & 1.2 & +0.05 & +1.7 & +0.03 \\
Belgium & 1.5 & 0.8 & 1.4 & -0.03 & -1.0 & -0.02 \\
Vietnam & 1.6 & 5.6 & 1.5 & +0.31 & +10.2 & +0.16 \\
Denmark & 2.8 & 0.9 & 1.6 & -0.05 & -1.5 & -0.04 \\
Romania & 1.1 & 3.0 & 1.2 & +0.20 & +6.7 & +0.08 \\
Indonesia & 3.6 & 4.2 & 1.7 & +0.24 & +8.0 & +0.28 \\
United Arab Emirates & 1.5 & 3.6 & 2.2 & +0.11 & +3.6 & +0.05 \\
Tunisia & 0.8 & 1.2 & 1.0 & -0.01 & -0.2 & 0.00 \\
Austria & 1.9 & 0.7 & 1.4 & -0.06 & -1.9 & -0.04 \\
Hungary & 1.6 & 2.4 & 1.3 & +0.13 & +4.0 & +0.06 \\
Malaysia & 1.2 & 3.4 & 2.0 & +0.12 & +3.8 & +0.05 \\
Finland & 2.3 & 1.4 & 1.6 & -0.01 & -0.3 & -0.01 \\
Morocco & 0.9 & 3.6 & 1.1 & +0.19 & +6.1 & +0.05 \\
Chile & 1.8 & 2.0 & 1.5 & +0.05 & +1.6 & +0.03 \\
\end{longtable}
}
\clearpage

\section{Discussion and conclusion}\label{sec:conclusion}

This paper measures the rate at which skilled emigrants return home and estimates how that rate responds to growth across many origins and destinations at once. Within a corridor, the developers who experienced more catch-up growth after leaving are the ones who returned, and the pull is projected to strengthen precisely where diasporas are largest, with return rising by roughly a tenth for India, Vietnam, and Nepal by 2035 under the IMF's growth paths.

Why should development policy care about a few percentage points of return? The first reason is demographic. Skilled migration moves people at the start of their most productive decades, so it shifts age structure as well as headcounts. For origins with young and growing workforces, India, Bangladesh, Vietnam, and Nepal among them, return converts a diaspora into mid-career talent at the moment their domestic technology sectors are best placed to absorb it. For origins already shrinking, for example in Eastern Europe, returnees offset an aging population. Destinations, meanwhile, face aging workforces of their own and have built innovation systems on the assumption that skilled immigrants, once arrived, mostly stay. Our projections suggest that this assumption will become less accurate on exactly the corridors that supply the most talent.

A second reason is that individual returns can generate significant value \citep{breschi2026geography}. A returning skilled migrant brings home more than they left with, because they carry back the skills, savings, and networks gained abroad \citep{dustmann2011}. The returnee-effects studies discussed above measure that gain: returnee managers raise patenting, returning scientists lift their colleagues' output, returnee directors raise firm value, and return raises development outcomes outright \citep{choudhury2016, fry2023, fryganguli2026, giannetti2015, bucheli2025}. Those studies take the number of returnees as given, and the aggregate gain from return depends on that number as well as on the effect per returnee. Our estimates show that the number responds to growth: the same catch-up that expands opportunity at home also brings more skilled migrants back.

A third reason is that circulation itself is valuable in creating knowledge spillovers. Skilled migrants who move back and forth carry tacit knowledge \citep{vdwouden2021} that is difficult to transmit at a distance \citep{vdwouden2023}, the know-how that helped build the technology industries of Taiwan and India \citep{saxenian2007}, and some of it is transmitted through diaspora ties even before anyone returns \citep{fackler2020}, ties that also steer migrants toward high-opportunity places to begin with \citep{ilyes2023}. For origin countries, the implication is that economic growth is a stronger driver of return than targeted policy. Returnee schemes and one-off incentives operate at the margin; the larger force is the income gap, which narrows as the economy develops and exerts a stronger pull as it does. This strengthens the case made by \citet{batista2025} for removing barriers to return and circulation. It is consistent, too, with the finding that returnees contribute most where the home investment climate and institutions are sound \citep{giannetti2015, bucheli2025}.

Our analysis has several limitations. Software developers are high-skilled labor whose work is often remote, and they maintain public GitHub profiles because those profiles carry signalhing value on the labor market; the developers we observe this way are a self-selected subset. We also note that while our identification is quite restrictive, estimates remain sensitive to time-varying corridor shocks correlated with growth. Our approach to measurement is also conservative and we suggest that the convergence coefficient is itself a lower bound, because we infer home from the first reported location which miscategorizes many developers who moved before the start of our dataset. Corrections described in the appendix raise the estimate by a third or more (Section~\ref{sec:lowerbound}). 

Despite these issues, our findings quantify the relationship between economic convergence and return migration. Given the importance of return migration for global brain circulation and knowledge flows, we suggest that this is an important first step in understanding the future of the geography of high-skilled labor and its crucial externalities. 

\section*{Data and ethics statement}

All data analyzed in this study come from public profiles on GitHub and Stack Overflow, obtained from the GHTorrent research archive and through the platforms' public interfaces in accordance with their terms of service. Locations are self-reported by users; we analyze them at the level of countries and corridors and report no individually identifying information. Aggregated corridor-level data and replication code will be made available upon publication.

\section*{Declaration of generative AI and AI-assisted technologies}

During the preparation of this work the authors used Anthropic's Claude to assist with data analysis and with drafting and editing the manuscript. Large language models also served as research tools in the construction and validation of the geocoding pipeline, as described in Appendix~\ref{app:pipeline}. After using these tools, the authors reviewed and edited the content and take full responsibility for the content of the publication.

\appendix

\section{Data pipeline and geocoding audit}\label{app:pipeline}

\subsection{From raw waves to the location panel}

This appendix describes how the raw profiles become the location panel, and audits the geocoding behind it. The profiles come from two successive sources. The waves from 2012 through 2021 are drawn from GHTorrent, the public research archive that mirrored GitHub's metadata for over a decade \citep{gousios2012, gousios2013}; the waves from 2024 through 2026 we collect ourselves from the GitHub REST API, which returns the same public profile fields. Each wave records the self-reported location string on every account's profile at the time of the data collection, and the steps below turn those strings into a panel of developers located by country.

The ten location waves union to a raw panel of 112.8 million person-wave observations for 44.3 million accounts; each wave contributes the accounts active in that year together with inactive accounts with a usable location. We remove 179,028 accounts identified as bots or organizations, combining the platform's account-type field, login-pattern rules (continuous-integration services and \texttt{-bot} suffixes, for example), activity thresholds implausible for a person, and published ground-truth bot lists. Where a profile lists no usable location in any wave, we fall back to the location on a linked Stack Overflow profile, geocoded through the same dictionary; this step never overrides a GitHub-derived location and provides 0.3\% of located developers and 0.15\% of located person-waves. Within a developer's history, waves without a report take the developer's nearest reported location, and such imputed rows are flagged; the audit below samples only directly reported strings. After these steps the analysis panel holds 42.6 million located person-waves for 10.0 million developers.

\subsection{Geocoding}

Free-text location strings resolve to countries through a curated dictionary of 1.2 million distinct strings, assembled with a large-language-model and web-search geocoder and extended by a conservative fallback that recovers a string absent from the dictionary only when at least three dictionary entries sharing its prefix agree on the country (about 5,200 strings recovered; 138,000 rejected as non-places, ambiguous, or insufficiently supported). Ambiguous names resolve to their modal usage among developers; \texttt{georgia}, for example, resolves to the country, which login-level evidence supports by roughly two to one, and genuinely unresolvable homonyms remain unassigned. Strings listing several places take the first listed one. Hong Kong and Macau are coded as their own territories throughout.

\subsection{A blind audit of geocoding quality}

We audited the dictionary on a blind sample of 1,000 distinct location strings drawn uniformly at random from the directly reported, geocoded strings in the panel (a frame of 795,046 distinct strings covering 32.2 million person-waves). A large language model, prompted with each raw string alone and blind to the pipeline's assignment, recorded the implied country; the pipeline's answer was joined only afterwards. The pipeline's assignment is correct for 97.6\% of strings and, because errors concentrate in rare strings, for 99.7\% of sampled users; the Wilson 95\% confidence interval for the string-level error rate is 1.6\% to 3.6\%. Every residual error is a non-place string, such as a joke or a fictional location, that inherited a country from the user's linked Stack Overflow profile; no real, resolvable place in the sample was assigned to the wrong country.

\subsection{Movers and returns}

A mover is a developer observed in at least two distinct countries across waves. Home is the first observed country, departure is the first wave observed outside it, and a return is the first subsequent wave observed at home again; onward moves to third countries do not censor the risk set. An independent recount of the location histories, which scans each developer's country sequence directly rather than passing through the panel construction, reproduces both headline counts exactly: 277,502 movers and 13,670 returns to the first observed country.

\section{Measurement corrections: home assignment and profile currency}\label{app:names}

Section~\ref{sec:lowerbound} reports two corrections to the headline convergence estimate; Table~\ref{tab:corrections} gives every sample restriction behind them. For the name-based cuts we predict each mover's nationality from the profile name with a statistical service (nationalize.io) that returns candidate countries with confidence scores, and keep movers whose assigned home is among the predicted countries, at successively stricter thresholds. The corroboration cut needs no names: we keep movers observed in the home country in at least two waves before departure, whose origin therefore cannot be a left-censoring artifact. For the profile-currency cut we keep movers who edited a non-location profile field, the display name, company, blog, or linked account, after departure; because the edited field is not the location field, this signal is independent of the return we measure.

\begin{table}[htbp]
\centering
\begin{tabular}{lcc}
\toprule
Sample & EC (SE) & $N$ \\
\midrule
Full risk set & $+0.315^{**}$ (0.098) & 490{,}756 \\
\addlinespace
\multicolumn{3}{l}{\emph{Home assignment}} \\
\quad Top-5 name match & $+0.424^{**}$ (0.155) & 190{,}785 \\
\quad Top-1 name match & $+0.018$ (0.194) & 110{,}729 \\
\quad Top-5, confidence $\geq 0.3$ & $+0.261^{+}$ (0.158) & 112{,}155 \\
\quad Home corroborated & $+0.301^{**}$ (0.108) & 423{,}313 \\
\addlinespace
\multicolumn{3}{l}{\emph{Profile currency}} \\
\quad Profile updated after departure & $+0.448^{*}$ (0.204) & 217{,}528 \\
\bottomrule
\end{tabular}
\caption{The convergence coefficient under measurement corrections. EC is the coefficient on cumulative economic convergence from the corridor specification of Table~\ref{tab:ecladder}, column~(4), re-estimated on each subsample with year and corridor fixed effects and two-way clustered standard errors (origin and destination country) in parentheses. The full risk set is the headline sample. The home-assignment cuts address origins misassigned by left-censoring; the profile-currency cut addresses returns hidden by stale locations. $^{+}\,p<0.1$, $^{*}\,p<0.05$, $^{**}\,p<0.01$, $^{***}\,p<0.001$.}
\label{tab:corrections}
\end{table}

The restricted samples are selected rather than random. The name prediction drops some correctly coded developers, those whose family name is common in a country other than their home, so the name-matched samples are not a random subset. The corroborated and profile-updated samples over-represent returnees, who tend to have longer pre-departure histories and better-maintained profiles. For both reasons we treat the cuts as evidence on the sign of the measurement bias rather than as preferred point estimates. Under our preferred filters, the top-5 name match and the profile update, the model estimates of the effect of convergence on return migration to be 30-40\% higher than in the overall dataset.

\section{Cultural distance does not moderate convergence}\label{app:interaction}

A natural conjecture is that catch-up growth pulls developers home less strongly on culturally distant corridors, where attachment to the origin has worn thinner. We test it by interacting cumulative convergence with standardized Kogut--Singh cultural distance on the horserace sample of Section~\ref{sec:robust}, using the same discrete-time logit, the same controls, and two-way clustering on origin and destination. The interaction is small and insignificant under both origin-and-destination fixed effects ($+0.046$, SE 0.135) and corridor fixed effects ($+0.166$, SE 0.155), and its point estimate runs, if anything, opposite to the conjecture; we therefore read the convergence effect as homogeneous across cultural distance: culture is associated with the level of return, and convergence with its change over time.

\section{Does GDP per capita proxy developer pay?}\label{app:salary}

\begin{figure}[htbp]
\centering
\includegraphics[width=0.78\linewidth]{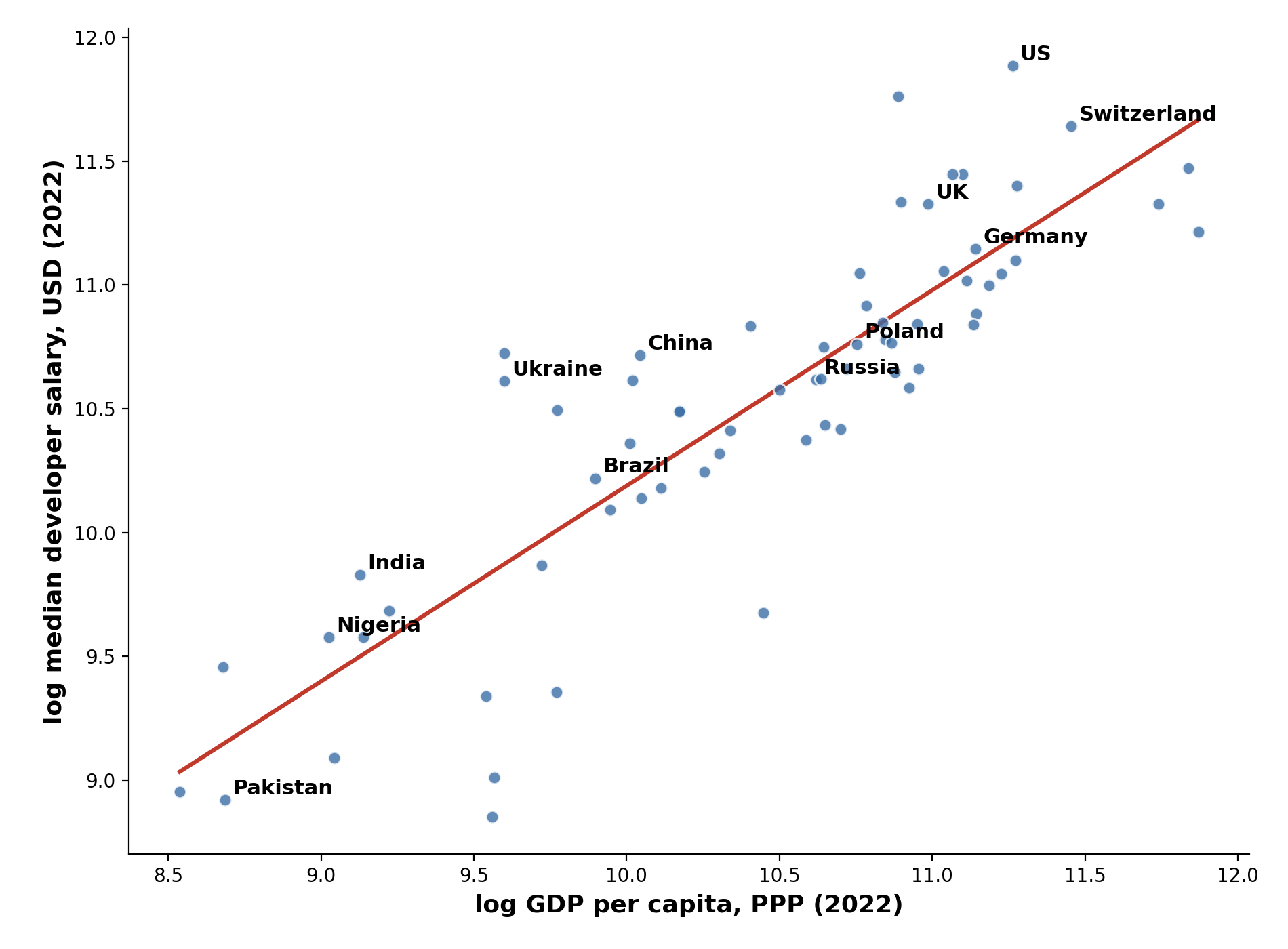}
\caption{Median developer salary against GDP per capita at PPP, by country, 2022 (Stack Overflow Developer Survey, countries with at least fifty salary respondents). The two track closely across the income range (Pearson $r = 0.88$); for example, developers in the United States earn about eight times what developers in India earn, close to the two countries' gap in GDP per capita. Each point is a country; the line is the ordinary-least-squares fit.}
\label{fig:salary}
\end{figure}

GDP per capita is an economy-wide average, so it is worth checking that it captures what software developers earn. We compare it to reported salaries from the Stack Overflow Developer Survey, pooled over 2017 through 2023, covering 282{,}917 developers across 76 countries. Across countries, median developer pay closely tracks GDP per capita at purchasing power parity, the measure we use in the paper (Figure~\ref{fig:salary}).

\section{Supplementary exhibits}\label{app:supp}

Table~\ref{tab:returnrates} gives the full return-rate ordering across all 53 origins with at least 1,000 movers. Tables~\ref{tab:distance} and~\ref{tab:cultural} report the distance and cultural-distance ladders referenced in Section~\ref{sec:robust}.

\begin{table}[htbp]
\centering
\begin{tabular}{lccccc}
\toprule
 & \multicolumn{5}{c}{DV: Return event} \\
\cmidrule(lr){2-6}
 & (1) & (2) & (3) & (4) & (5) \\
\midrule
Log distance (km) & $-0.076$ &  & $-0.028$ & $0.064$ & $0.036$ \\
 & (0.061) &  & (0.063) & (0.045) & (0.040) \\
\addlinespace
Timezone difference (hours) &  & $-0.024$ & $-0.017$ & $-0.038^{+}$ & $-0.024$ \\
 &  & (0.023) & (0.032) & (0.021) & (0.016) \\
\addlinespace
Years since departure & $-0.107^{***}$ & $-0.107^{***}$ & $-0.107^{***}$ & $-0.128^{***}$ & $-0.126^{***}$ \\
 & (0.012) & (0.012) & (0.012) & (0.010) & (0.010) \\
\midrule
Year FE & \checkmark & \checkmark & \checkmark & \checkmark & \checkmark \\
Origin country FE & & & & \checkmark & \checkmark \\
Destination country FE & & & & & \checkmark \\
\midrule
Observations & 535,506 & 535,506 & 535,506 & 535,504 & 535,502 \\
Base event rate & 0.0250 & 0.0250 & 0.0250 & 0.0250 & 0.0250 \\
\bottomrule
\end{tabular}
\caption{Geographic distance, time zones, and the return hazard. Discrete-time logit hazard models on the post-departure risk set, consistent sample across columns. Two-way clustered standard errors (origin and destination country) in parentheses. $^{+}\,p<0.1$, $^{*}\,p<0.05$, $^{**}\,p<0.01$, $^{***}\,p<0.001$.}
\label{tab:distance}
\end{table}

\begin{table}[htbp]
\centering
\begin{tabular}{lccc}
\toprule
 & \multicolumn{3}{c}{DV: Return event} \\
\cmidrule(lr){2-4}
 & (1) & (2) & (3) \\
\midrule
Cultural distance (Kogut--Singh) & $-0.070^{**}$ & $-0.050^{*}$ & $-0.069^{***}$ \\
 & (0.026) & (0.020) & (0.015) \\
\addlinespace
Log distance (km) & $-0.047$ & $-0.019$ & $0.004$ \\
 & (0.060) & (0.030) & (0.027) \\
\addlinespace
Years since departure & $-0.107^{***}$ & $-0.126^{***}$ & $-0.125^{***}$ \\
 & (0.012) & (0.010) & (0.010) \\
\midrule
Year FE & \checkmark & \checkmark & \checkmark \\
Origin country FE & & \checkmark & \checkmark \\
Destination country FE & & & \checkmark \\
\midrule
Observations & 524,024 & 524,023 & 524,023 \\
Base event rate & 0.0250 & 0.0250 & 0.0250 \\
\bottomrule
\end{tabular}
\caption{Cultural distance and the return hazard. Discrete-time logit hazard models on the post-departure risk set, consistent sample across columns. Two-way clustered standard errors (origin and destination country) in parentheses. $^{+}\,p<0.1$, $^{*}\,p<0.05$, $^{**}\,p<0.01$, $^{***}\,p<0.001$.}
\label{tab:cultural}
\end{table}

\clearpage
{\scriptsize\setlength{\tabcolsep}{6pt}\renewcommand{\arraystretch}{0.82}
\begin{longtable}{lrrrrr}
\toprule
Origin & Movers & Returned & Return rate (\%) & Movers (5-yr) & Return rate (5-yr, \%) \\
\midrule
\endfirsthead
\toprule
Origin & Movers & Returned & Return rate (\%) & Movers (5-yr) & Return rate (5-yr, \%) \\
\midrule
\endhead
\bottomrule
\caption{Return rates by origin country, for the 53 origins with at least 1{,}000 movers, sorted by the five-year rate. The pooled rate counts all returnees regardless of how long each has been observed and is biased down by recent departures; the five-year rate is the share returning within five years among the 2016, 2019, and 2021 departure cohorts, those an observed wave follows exactly five years later.}\label{tab:returnrates}\\
\endlastfoot
Spain & 4,662 & 553 & 11.86 & 1,419 & 16.1 \\
France & 7,895 & 858 & 10.87 & 2,360 & 15.0 \\
Poland & 2,806 & 280 & 9.98 & 754 & 14.9 \\
Australia & 4,419 & 442 & 10.00 & 1,412 & 14.1 \\
Finland & 1,108 & 110 & 9.93 & 322 & 14.0 \\
Japan & 2,974 & 285 & 9.58 & 768 & 13.5 \\
Sweden & 2,534 & 219 & 8.64 & 725 & 13.5 \\
Indonesia & 1,337 & 84 & 6.28 & 240 & 12.9 \\
United States & 22,260 & 1,791 & 8.05 & 6,063 & 12.9 \\
Denmark & 1,347 & 116 & 8.61 & 373 & 12.9 \\
Germany & 9,652 & 717 & 7.43 & 2,381 & 12.9 \\
South Korea & 2,484 & 153 & 6.16 & 495 & 12.5 \\
Canada & 9,402 & 645 & 6.86 & 2,463 & 12.4 \\
Colombia & 1,743 & 86 & 4.93 & 368 & 12.0 \\
Portugal & 2,131 & 177 & 8.31 & 564 & 11.5 \\
Netherlands & 3,831 & 246 & 6.42 & 1,004 & 11.5 \\
Hungary & 1,209 & 70 & 5.79 & 314 & 11.1 \\
New Zealand & 1,860 & 122 & 6.56 & 552 & 11.1 \\
Greece & 1,537 & 104 & 6.77 & 434 & 10.8 \\
Austria & 1,237 & 85 & 6.87 & 324 & 10.8 \\
Switzerland & 2,523 & 157 & 6.22 & 696 & 10.8 \\
Romania & 1,340 & 93 & 6.94 & 423 & 10.2 \\
Italy & 4,622 & 324 & 7.01 & 1,272 & 9.8 \\
Brazil & 12,474 & 812 & 6.51 & 3,536 & 9.4 \\
Belgium & 1,475 & 93 & 6.31 & 416 & 9.1 \\
Taiwan & 1,853 & 100 & 5.40 & 497 & 8.7 \\
United Kingdom & 14,503 & 763 & 5.26 & 4,223 & 7.6 \\
Chile & 1,023 & 49 & 4.79 & 238 & 7.6 \\
Ireland & 2,366 & 103 & 4.35 & 670 & 7.0 \\
China & 13,751 & 534 & 3.88 & 3,176 & 6.9 \\
Mexico & 2,283 & 107 & 4.69 & 620 & 6.6 \\
Singapore & 2,568 & 89 & 3.47 & 602 & 6.0 \\
Vietnam & 1,463 & 54 & 3.69 & 337 & 5.9 \\
Ukraine & 7,252 & 164 & 2.26 & 1,309 & 5.5 \\
Turkey & 5,796 & 168 & 2.90 & 989 & 5.1 \\
Hong Kong & 1,886 & 47 & 2.49 & 420 & 4.8 \\
Sri Lanka & 1,516 & 29 & 1.91 & 217 & 4.6 \\
India & 34,798 & 860 & 2.47 & 6,819 & 4.6 \\
Malaysia & 1,202 & 30 & 2.50 & 274 & 4.4 \\
Iran & 3,752 & 79 & 2.11 & 651 & 4.3 \\
Argentina & 2,825 & 96 & 3.40 & 636 & 4.2 \\
Bangladesh & 3,393 & 60 & 1.77 & 406 & 4.2 \\
United Arab Emirates & 1,320 & 30 & 2.27 & 223 & 4.0 \\
Morocco & 1,065 & 14 & 1.31 & 201 & 4.0 \\
Pakistan & 4,063 & 80 & 1.97 & 469 & 3.8 \\
South Africa & 1,858 & 42 & 2.26 & 470 & 3.8 \\
Russia & 18,453 & 373 & 2.02 & 2,273 & 3.6 \\
Belarus & 4,470 & 47 & 1.05 & 310 & 3.2 \\
Nepal & 1,576 & 14 & 0.89 & 250 & 3.2 \\
Egypt & 2,206 & 43 & 1.95 & 440 & 3.2 \\
Nigeria & 3,004 & 44 & 1.46 & 246 & 2.4 \\
Venezuela & 2,329 & 25 & 1.07 & 845 & 1.1 \\
Tunisia & 1,248 & 10 & 0.80 & 262 & 0.8 \\
\end{longtable}
}

\clearpage
\bibliography{bibliography}

\end{document}